\documentclass[]{aiaa-tc}

\usepackage{lettrine}
\usepackage{url}
\usepackage{pdfpages}
\usepackage{multienum}
\usepackage{multicol}
\usepackage{fancyhdr}
\usepackage{amsmath}
\usepackage{multirow} 
\usepackage{booktabs} 
\usepackage{amssymb}
\usepackage{amsfonts}
\usepackage{bm}
\usepackage{caption}
\usepackage{subcaption}
\usepackage{comment}
\usepackage{amssymb}
\usepackage{pifont}
\usepackage[utf8]{inputenc}
\usepackage{overpic}
\usepackage{graphicx}
\graphicspath{{images/}}
\usepackage{parskip}
\usepackage{float}
\usepackage[official]{eurosym}
\usepackage{tabto}
\usepackage{siunitx}
\usepackage{tikz}
\usetikzlibrary{arrows.meta,calc,decorations.markings,math,arrows.meta,shadows}
\usepackage{lscape}
\usepackage{hyperref}
\newcommand{\argmin}[2]{\underset{#1}{\mathrm{argmin}}\,{#2}}
\newcommand{\triad}[2]{$\left (#1 \rightleftharpoons #2\right )$}

\title{Data-driven analysis of anomalous transport 
and three-wave-coupling effects 
in $E \times B$ plasma discharges}

\author{
Borja Bayón-Buján\thanks{Research Assistant, bbayon@pa.uc3m.es}~%
~, Enrique Bello-Benítez\thanks{PhD Candidate, ebello@ing.uc3m.es}~%
~, Jiewei Zhou\thanks{Assistant Professor, jzhou@pa.uc3m.es}
~and Mario Merino\thanks{Professor, marmerin@ing.uc3m.es}\\%
{\normalsize\it{Department of Aerospace Engineering, Universidad Carlos III de Madrid, Leganés, Spain.}}\\%
\\%
}

\begin{document}

\maketitle

\begin{abstract}
Collisionless cross-field electron transport in an $E\times B$ configuration relevant for electric propulsion is studied using data from a ($z,\theta$) full-PIC simulation.
Higher-order spectral analysis shows that transport is dominated by the in-phase interaction of the oscillations of the azimuthal electric field and the electron density
associated to the first electron cyclotron drift instability (ECDI) mode. 
A secondary contribution emanates from a lower-frequency mode, not predicted by linear ECDI theory, while higher modes have a minor direct impact on transport. 
However, a bicoherence analysis reveals that strong phase couplings exist among the ECDI modes, and
a sparse symbolic regression spectral model,
based on the three-wave coupling equations, suggests an inverse energy cascade as the most likely explanation, thus  suggesting that higher modes contribute indirectly to transport by quadratic power transfer to the first mode.
This work provides new insights into the dynamics of anomalous plasma transport in $E\times B$ sources and the underlying processes governing energy distribution across different scales, and supports the validity of weak turbulence theory to examine their behavior.
\end{abstract}

\section{Introduction}

Anomalous cross-field electron transport in $\bm E \times \bm B$ plasmas remains a key unexplained phenomenon that drives the performance losses of many devices. In the field of electric propulsion, anomalous transport is well-known to occur in Hall thrusters \cite{kaga20a,boeu17}, but is likely present too in other devices such as electrodeless plasma thrusters \cite{hepn20b,taka22a,madd24b}. 
The significant electron drift in the azimuthal direction of Hall thrusters is known to give rise to various azimuthal oscillations and instabilities, potential explanations for the observed anomalous transport. It is generally agreed that cross-field anomalous transport of electrons occurs mainly due to the $\overline{n_{e} E_{y}}$ time-averaged term in the azimuthal momentum equation \cite{jane66,bell22c,lafl16a}.
However, the underlying mechanisms giving rise to these azimuthal oscillations are still a subject of active research. 
The electron cyclotron drift instability (ECDI) has been identified as one of the probable actors behind the anomalous electron transport. When ion acoustic modes align with electron cyclotron resonances $m\omega_{ce}$ due to Doppler shift from electron $\bm E \times \bm B$ motion, instability occurs. 
Power transfers from collective electron motion to ions until the instability enters the nonlinear regime and saturation happens.
Even though the linear theory of the ECDI is established\cite{fors70,wong70,ducr06,cava13}, the exact mechanisms or transport laws and the energy balance in the nonlinear regime are still a matter of discussion. 
Some points that need to be further clarified are the suggested development of an inverse-cascade process\cite{janh18, smol23}, the transition of the ECDI to an ion-acoustic mode\cite{cava13, lafl17b,tacc19c}, and the nonlinear effects on transport of the coexistence of several active modes \cite{janh18, hara20, mike20, brow23}  

On the computational side, Particle-in-cell (PIC) codes have been widely used to simulate the effect of instability-induced oscillation on Hall plasmas; either in canonical\cite{bell24,janh18,lafl16a,tacc19c,smol23,chen23} or more realistic/applied\cite{char19b,adam04,lafl17b} scenarios. 
Recently, data-driven techniques have been adapted and tested on electric propulsion plasmas, as a valuable complement to  the researcher toolbox. These techniques also offer the advantage of being flexible in their applicability. Examples include the use data coming from axial-radial simulations of a Hall discharge to identify and isolate dominant dynamic regimes through Proper Orthogonal Decomposition (POD) and Dynamic Mode Decomposition (DMD) \cite{madd22a, pera23a,  fara23one}; further work used one of these datasets together with the Sparse Identification of Nonlinear Dynamics (SINDy) algorithm \cite{bayo24a} to obtain parsimonious equations (both physically-meaningful and with very few terms) of the breathing mode dynamics directly from the time-series of plasma variables. Previously, symbolic regression  had also been proposed to obtain data-driven closures to the anomalous transport problem \cite{jorn18}.

Weakly-nonlinear plasma theory treats a perturbation as a superposition of (linear) eigenmodes whose amplitude can vary in time due to nonlinear interactions, which can be divided in two main types: wave-wave and wave-particle interactions \cite{sagd69, weil77}. Higher-order spectral analysis of the macroscopic fields can be used 
to interrogate the data for possible wave-wave interactions; particularly, the bispectrum and the bicoherence have been used successfully to identify quadratic nonlinear interactions in space plasmas \cite{lago89} , fusion plasmas \cite{naga06, mill08},  and very recently in electric propulsion in diverse configurations \cite{yama23b,brow23,madd24b}. Among the latter, a recent study\cite{yama23b} used this technique to relate low and high-frequency density oscillations in a Hall discharge from microwave interferometry measurements, while another used the bispectrum together with three-wave coupling theory to obtain growth rates and nonlinear interaction coefficients from experimental data \cite{brow23} based on the Kim-Ritz method \cite{kim96, ritz89}, identifying the relevant spectral components for instability, albeit based on an mathematically overdetermined problem.

In this paper, we analyze full-PIC simulation data from Bello-Benítez et al. \cite{bell24, bell22c} 
to first dissect the contribution of the different oscillatory modes to the cross-field electron current $j_{ze}$, by examining the frequency spectrum 
of the  $\overline{n_{e} E_{y}}$ term and the magnitude and phase difference of the $n_e$ and $E_y$ oscillations. Next, we investigate the existence of nonlinear couplings between the $n_e$, $E_y$, and $j_{ze}$ time signals through a mutual bicoherence analysis. 
The analysis allows us to 
identify the frequency bands primarily responsible for transport and 
to confirm
that the dominant mechanism for cross-field electron transport is the in-phase density-field fluctuations related to the first mode of the ECDI.

However, this result by itself, does not provide a complete understanding of the
source of the energy in the fluctuation, nor its evolution from the unstable frequencies to the rest of the spectrum by means of nonlinear couplings.
Therefore, in the second part of this work we build a reduced spectral model for the nonlinear energy coupling in the $E_y$ spectrum by means of sparse-regression data-driven modeling of energy evolution at dominant frequencies. Our study demonstrates that quadratic, three-wave coupling can effectively explain at least part of the energy transfer from instability frequencies to the bands responsible for cross-field transport by means of an inverse energy cascade.

The rest of the paper is structured as follows: Section \ref{sec:simoverview} provides a brief overview of the simulation and the data used to carry out this work, Section \ref{sec:methods} introduces the techniques used to analyze and model the data, Section \ref{sec:results} outlines the results obtained, divided into Section \ref{sec:spectralanalysis} for the identification of the different modes present in the discharge, Section \ref{sec:analysis} for the quantification of dominant contributions to the anomalous electron current, and \ref{sec:modeling} for the reduced models of power transfer based on nonlinear three-wave coupling. Finally, Section \ref{sec:conclusions} gathers the conclusions of the study as well as future steps.
The reader is adviced to check reference \cite{bell24} before continuing, as the code, simulations, and fundamental structure of the observed fluctuations, are described there.
A preliminary version of this work was presented in a recent conference \cite{bayo24b}.

\section{Simulation overview}
\label{sec:simoverview}

The simulation of a canonical annular $\bm{E} \times \bm{B}$ plasma discharge, representative of a Hall thruster channel, is used in this work.  
The simulation domain corresponds to the 
axial-azimuthal midchannel surface of the discharge.
The simulation is carried out with the in-house, 2D, electrostatic full-PIC code named PICASO\cite{bell22c,bell24,mari24a,bell24b}.
The details of the code and the simulation are given in those references.
In essence, the equations of motion of the ion and electron macroparticles are solved with an explicit, momentum-conserving Boris algorithm  and the interpolation and weighting schemes implement first-order bi-linear shape functions.
The code is implemented in Fortran90 and the operations on macroparticles are parallelized following a particle-decomposition strategy using 
shared-memory OpenMP.
The Poisson solver used here employs second-order finite-differences to discretize the Laplace operator and  PARDISO Intel MKL direct solver to invert the resulting linear system. 
 
\begin{figure}
    \centering
    \includegraphics[width=0.7\textwidth]{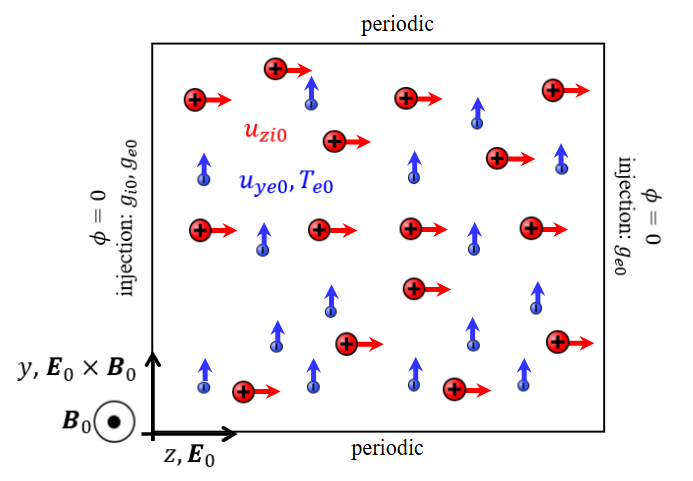}
    \caption{Diagram sketching the full-PIC simulation domain, boundary conditions and initial equilibrium state.}
    \label{fig:sim_setup}
\end{figure}

The simulation settings are summarized in Figure \ref{fig:sim_setup} and in Table \ref{tab:params}.
A collisionless plasma composed of electrons and hydrogen ions is considered. Because it has been shown that the entire spectrum scales as $1/\sqrt{m_i}$, the choice of $m_i$ is arbitrary in terms of the conclusions of spectral analysis. Thus, the present analysis can be generalized for heavier species without loss of generality. The remaining settings aim to replicate essential aspects of the physics needed to trigger and sustain the ECDI\cite{bell22c,bell24}, while excluding other factors like field inhomogeneities, ionization, and collisions, as noted in prior studies. An additional simulation, run for 30 $\mu s$ with a 10× higher sample rate, is included in the final part of this study to achieve finer spectral resolution and validate the shorter simulation's results.

A fixed magnetic field $\bm B_0 = B_0 \bm 1_x$ is considered in the out-of-plane direction, together with a perpendicular equilibrium electric field $\mathcal{\bm E} = E_0 \bm 1_z$ in the axial direction. Here and throughout the paper, the sub-index `0' stands for equilibrium conditions.
As usually assumed in Hall thruster plasmas, ions are considered unmagnetized and to be in an unaccelerated equilibrium state (i.e., they do not feel $\bm B_0$ nor $\bm E_0$). 
Electron equilibrium, on the other hand, is defined by the  $\bm E_0\times \bm B_0$ drift.
Nevertheless, both species do respond to the perturbation electric field, $\bm E = -\nabla \phi$, obtained from the solution to Poisson equation. 
This approach, while it implies lack of a consistent equilibrium for ions and exact energy conservation, has been used successfully in previous works to examine the development of instabilities in $\bm E\times\bm B$ plasmas\cite{janh18,smol23,bell24}. 

The plasma simulation is initialized in homogeneous equilibrium state with cold ions drifting with homogeneous velocity $\bm u_{i0} = u_{zi0} \bm 1_z$  and Maxwellian electrons with temperature $T_{e0}$ and mean velocity $\bm u_{e0} = u_{ye0} \bm 1_y$, corresponding to the drift $u_{ye0} = E_0/B_0$. The density of both species is $n_0$.
The initial macroparticle populations of electrons and ions are randomly generated with these properties. 

On the $y$ boundaries, periodic boundary conditions are imposed on the particles and on the potential. Then, direction $y$ represents the azimuthal direction in a Hall thruster.
Particles reaching axial $z$ boundaries are removed from the simulation, but there is continuous injection of ions through the left boundary with  flux density $n_0 u_{zi0}$ and of electrons through left/right boundaries with fluxes $\pm n_0 c_{e0}/\sqrt{2 \pi}$, corresponding to equilibrium conditions. The perturbation potential $\phi$ on the $z$ boundaries is set to zero.
Because the discrete nature of the simulation domain in the $y$ direction, the possible wavenumbers are limited to integer multiples of $L_y$. Thus, this parameter is purposely chosen such that for each of the first ECDI resonances, $k_{}L_y=1+6m$ ($m=1,2,3,4$) approximately match the peaks of linear growth rate, as given by linear theory.
 
\begin{table}[]
\centering
\begin{tabular}{|c|c|}
\hline
\textbf{Description and symbol}     & \textbf{Value and units} \\ \hline     
Ion mass, $m_i$                     & 1 amu                      \\ \hline
Applied electric field, $E_0$       & $10^4$ V/m               \\ \hline
Applied magnetic field, $B_0$       & 200 G                    \\ \hline
Plasma density, $n_0$               & $10^{17}$ m$^{-3}$       \\ \hline
Ion axial velocity, $u_{zi0}$       & 10 km/s                  \\ \hline
Electron temperature, $T_{e0}$      & 6 eV                     \\ \hline
Azimuthal domain length, $L_y$    & 5.359 mm                 \\ \hline
Axial domain length, $L_z$          & 2.679 mm                 \\ \hline\hline
$E_0\times B_0$ drift, $u_{ye0}$        & 500 km/s                 \\ \hline
Electron thermal speed, $c_{e0}$    & 1027 km/s                \\ \hline
Ion sound speed, $c_{s0}$           & 23.97 km/s               \\ \hline
Debye length, $\lambda_{D0}$        & 57.58 $\mu$m             \\ \hline
Electron Larmor radius, $\rho_{e0}$ & 292.0 $\mu$m             \\ \hline
Electron plasma frequency, $\omega_{pe0}$ & 2.839 GHz         \\ \hline
Electron gyrofrequency, $\omega_{ce}$     & 0.5600 GHz        \\ \hline
Ion plasma frequency, $\omega_{pi0}$      & 66.26 MHz         \\ \hline
Lower-hybrid frequency, $\omega_{lh}$     & 13.07 MHz         \\ \hline\hline
Number of cells in $y$ direction, $N_y$    & 100              \\ \hline
Number of cells in $z$ direction, $N_z$    & 50               \\ \hline
Number of particles per cell, $N_{\textmd{ppc}}$ & $\sim 200$ \\ \hline
Time step, $\Delta t$                    & $5 \times 10^{-12}$ s \\ \hline
Number of time steps, $N_t$              & $10^6$             \\ \hline
Cell size, $\Delta y$, $\Delta z$        & 53.59 $\mu$m       \\ \hline
\end{tabular}
\caption{Physical and numerical parameters of the reference simulation case. The subscript `0' stands for initial equilibrium conditions. Derived parameter values are included for completeness.}
\label{tab:params}
\end{table}


\section{Methods}
\label{sec:methods}

In the following, we introduce the spectral quantities and the formalism behind the sparse identification of nonlinear dynamics (SINDy) technique, employed in the reminder of the paper.

\subsection{High order spectral analysis}

A signal sampled over a spatiotemporal domain may be described in terms of a collection of discrete Fourier modes:
\begin{equation}
x(t_i, y_j)=\sum_m \hat{x}(\omega_m, y_j)e^{-\text{i}\omega_mt_i} =\sum_n \tilde{x}(k_n, t)e^{\text{i} k_ny_j}\\
=\sum_n\sum_m \hat{\tilde{x}}(\omega_m, k_n)e^{-\text{i}\omega_mt_i+\text{i} k_n y_j }
\label{eq:fourier}
\end{equation}
where $\text{i}$ is the imaginary unit and $\omega_m$ and $k_n$ cover both positive and negative frequencies with the complex conjugate property that $\hat{x}^{\ast}(-\omega_m, y_j)=\hat{x}(\omega_m, y_j)$, $\hat{\tilde{x}}^{\ast}(-\omega_m, -k_n)=\hat{\tilde{x}}(\omega_m, k_n)$ and so on.

The Power Spectral Density (PSD) is defined as
\begin{align}
P_{ab}(\omega,k_{}) = \left \langle\hat{\tilde{x}}_a(\omega, k_{})\hat{\tilde{x}}^{\ast}_b(\omega, k_{})\right \rangle.
\end{align}
Here, \(\left \langle \cdot \right \rangle\) denotes averaging over multiple realizations, and \(*\) indicates complex conjugation. Summing over the $y$ direction we can also define
\begin{align}
P_{ab}(\omega) = \frac{1}{N_j}\sum_j\left \langle\hat{{x}}_a(\omega, y_j)\hat{{x}}^{\ast}_b(\omega, y_j)\right \rangle,
\end{align}
where $N_j$ denotes the number of points, which is related to $P_{ab}(f,k)$ through
\begin{align}
P_{ab}(\omega) = \frac{1}{N_j}\sum_j\left \langle\sum_n\hat{\tilde{x}}_a(\omega, k_n)e^{\text{i} k_n y_j}\sum_m\hat{\tilde{x}}^{\ast}_b(\omega, k_m)e^{\text{i} k_m y_j}\right \rangle =
\sum_n P_{ab}(\omega,k_n).
\label{eq:demo}
\end{align}
Here, we have used $\hat{x}(\omega, y_j) = \sum_n \hat{\tilde{x}}(\omega, k_n) e^{\text{i} k_n y_j}$ and the orthogonality properties of the discrete Fourier transform.
We can similarly define $P_{ab}(k)$ summing over time $t$.

When nonlinear behavior is present, interactions between different modes can take place. In the simplest case of wave-wave coupling, three-wave coupling due to quadratic terms can occur if the resonance conditions,
\begin{align}
\omega_1+\omega_2&=\omega_3,
&
k_1 + k_2 &=k_3,
\label{eq:resonanceconditions}
\end{align}
are satisfied among three propagating modes ($\omega_1$, $k_1$), ($\omega_2$, $k_2$) and ($\omega_3$, $k_3$). Here the analysis will be done along a single spatial dimension, but analogous conditions apply to the remaining others.
Given three signals $x_j(t, y)$ for $j=a,b,c$ , we define their bispectrum
as the  third-order cumulant spectrum \cite{kim79},
\begin{equation} 
\mathfrak{B}_{abc}(\omega_1, \omega_2, k_1, k_2) =  \left \langle \hat{\tilde{x}}_a(\omega_1, k_1)\hat{\tilde{x}}_b(\omega_2, k_2)\hat{\tilde{x}}^*_c(\omega_1+\omega_2, k_1+k_2) \right \rangle.
\label{eq:bispectrum}
\end{equation}
Reduced versions of the bispectrum can be defined
in terms of only the frequencies or the wavenumbers alone. For example, in terms of frequencies only we define, by averaging over the $y$ direction, 
\begin{equation}
\mathfrak{B}_{abc}(\omega_1, \omega_2) =  \frac{1}{N_j}\sum_j \left \langle \hat{x}_a(\omega_1, y_j)
\hat{x}_b(\omega_2, y_j)\hat{x}^*_c(\omega_1+\omega_2,y_j) \right \rangle
=\sum_n\sum_l\mathfrak{B}_{abc}(\omega_1, \omega_2, k_n, k_l).
\end{equation}
The latter relation can be proven with an expansion similar to the one in equation \eqref{eq:demo}.  
And similarly, $\mathfrak{B}_{abc}(k_1, k_2)$ can be computed from $\mathfrak{B}_{abc}(\omega_n, \omega_l, k_1, k_2)$, summing over time $t$.

From the bispectrum, the bicoherence is defined as
\begin{equation} \label{eq:bicoherence}
b_{abc}(\omega_1, \omega_2, k_1, k_2) = \frac{\left| \mathfrak{B}_{abc}(\omega_1, \omega_2, k_1, k_2) \right|}{\sqrt{\left \langle \left | \hat{\tilde{x}}_a(\omega_1, k_1) \hat{\tilde{x}}_b(\omega_2, k_2) \right |^2 \right \rangle\left \langle \left | \hat{\tilde{x}}^*_c(\omega_1+\omega_2, k_1+k_2) \right |^2 \right \rangle}}.
\end{equation}
When \(x_a\), \(x_b\), and \(x_c\) are the same signal, one speaks of $b_{aaa}\equiv b_{a}$ as the (self-)bicoherence of that signal; otherwise, the term cross-bicoherence (among different signals) is used.
The self-bicoherence features a number of symmetries that can be exploited to speed up its computation. Equivalently to what has been shown with the bispectrum, $b_{abc}(\omega_1, \omega_2)$ and $b_{abc}(k_1, k_2)$ can be defined.

  
A value of $b=1$ indicates perfectly phase-locked modes, suggesting three-wave coupling, whereas random phases or noise will lead to bicoherence values closer to $0$ as the number of realizations or the noise level increases; Specifically, the 95 percent significance level for null bicoherence computed over $N$ realizations is approximately $\sqrt{3/N}$\cite{elga88}.  In our case the minimum number of realizations used is $N=1300$, corresponding to significant bicoherence above $b=0.05$.
Discrete interactions between two modes show up as ``islands" of high bicoherence, whose width corresponds to the spectral broadening of interacting peaks.  
Continuous interactions of a single frequency with a broader band are shown as lines or segments, either verticals ($\omega_1=\text{const}$), horizontals ($\omega_2=\text{const}$) or diagonals ($\omega_3=\text{const}$). 
Note that the bicoherence is high when quadratic phase coupling exists among modes. However, the bicoherence itself does not discriminate the direction of power flow among them, and this directionality needs to be  studied by other methods, such as by fitting the underlying three-wave coupling equations.
 
\subsection{Sparse Identification of Nonlinear Dynamics (SINDy)}
\label{sec:methods_SINDy}

As explained in \cite{bayo24a}, the basic form the SINDy framework \cite{brun16b} goes as follows. We consider a dynamical system given by a state vector $\bm{x}(t) = \left[x_1(t),x_2(t),\ldots,x_I(t)\right]^T$ in a state space $\mathbb X$, governed by a set of ordinary differential equations of the form
\begin{equation}
    \dot{x}_i(t)=f_i(\bm{x},t),
    \label{eq:dyneq}
\end{equation}
where $f_i$ ($i=1,\ldots,I$) are unknown functions of the state, and possibly, time. We aim to write write each $f_i$ in \eqref{eq:dyneq} as 
\begin{equation}
f_i(\bm{x},t) = \beta_{ij} \Theta_j (\bm{x},t),
\label{eq:dyneq2}
\end{equation}
where $\Theta_j$ ($j=1,\ldots , J$) is a chosen collection of functions (termed ``features'') and  $\beta_{ij}$ a (sparse) array of coefficients to be determined.

If a realization of the dynamical system has data $x_i(t_k) \equiv \hat x_{ik}$ at discrete time instants $t_k$ ($k=0,\ldots, K$), potentially subject to noise, it is possible to estimate the coefficients $\beta_{ij}$ from the following linear system of equations:
\begin{equation}
\label{eq:reg}
\dot{\hat{x}}_{ik} = \beta_{ij}\hat\Theta_{jk}
\end{equation}
where $\dot{\hat{x}}_{ik}$ is a numerical estimate of the time derivatives of the state from the data, e.g. using finite differences, and $\hat\Theta_{jk}\equiv\Theta_j(\hat{\bm x}(t_k), t_k)$. 

This set of equations is typically strongly overdetermined, as we have many more equations than unknown coefficients, $K \gg IJ$. 
Naively solving for $\beta_{ij}$ by minimization of the least-square error
\begin{equation}
\label{eq:errLS}
    \varepsilon^{S}= \frac{1}{N}\frac{1}{\hat{\sigma}^2_{\dot{x}}}\sum_{i,k} \left ( \dot{\hat{x}}_{ik} - \beta_{ij}\hat{\Theta}_{jk} \right )^2,
\end{equation}
where $N$ stands for the sample size and $\hat{\sigma}^2_{\dot{x}}$ for the variance of the numerical derivatives, typically yields a full $\beta_{ij}$ matrix with 
most coefficients  different from zero. This is usually undesired, as the resulting models exhibit an unaffordable complexity and lack simple physical interpretations.

What SINDy proposes is finding $\beta_{ij}$ through the minimization of the sum of a Least Square error $\varepsilon^{S}$, plus a
sparsity-promoting regularization term (or penalty), $\varepsilon^\lambda$,
\begin{equation}
\beta_{ij} = \argmin{\beta_{ij}}{\left(\varepsilon^{S}+\varepsilon^\lambda\right)}.
\label{eq:min1}
\end{equation}
By regularizing to promote sparsity in the solution $\beta_{ij}$, the algorithm is shown to regress on the features most relevant to the dynamics and discard the rest \cite{brun16b}. In the present work, we use the ALASSO penalty \cite{cort21, zou06} 
\begin{equation}
\label{eq:errreg}
    \varepsilon^{\lambda}=\left | a_{ij}\beta_{ij} \right |\; \; \; \text{with }a_{ij}=\frac{\lambda_i}{\beta_{ij}^*},
\end{equation}
where the $\lambda_i$ are hyperparameters which set the relative weight of the regularization term over the error term for each state variable, and the term-specific weights $\beta_{ij}^*$ are the coefficient estimates coming from optimizing $\varepsilon^S$ alone, 
\begin{equation}
\beta_{ij}^* =  \argmin{\beta_{ij}}{\left(\varepsilon^{S}\right)}.
\end{equation}
This form of $\varepsilon^\lambda$ puts a large penalty on small coefficients while reducing biases on the larger coefficients. 
The ALASSO penalty also has the benefit of leading to a computationally-efficient convex minimization problem, and offers consistent variable selection and correct coefficient estimation as the number of samples $K$ tends to infinity (given that all relevant features are included in the chosen function library, and available data spans the whole state space sufficiently) \cite{zou06}.

For each state variable $x_i$ sweeping the regularization parameter $\lambda_i$ from 0 to infinity and plotting the error and complexity results in an L-shaped Pareto front, from which the optimal model can be selected based on the knee inflexion point \cite{hans92,bayo23} and previous knowledge of the system. Finally, the accuracy of the model can be evaluated from its $R^2$ score, defined from \eqref{eq:errLS} as
\begin{equation}
    \label{eq:R2score}
    R^2=1-\varepsilon^S
\end{equation}

\section{Results and discussion}
\label{sec:results}
\subsection{Spectral analysis of the discharge}
\label{sec:spectralanalysis}
\begin{figure}
    \centering
    \includegraphics[width=1\textwidth]{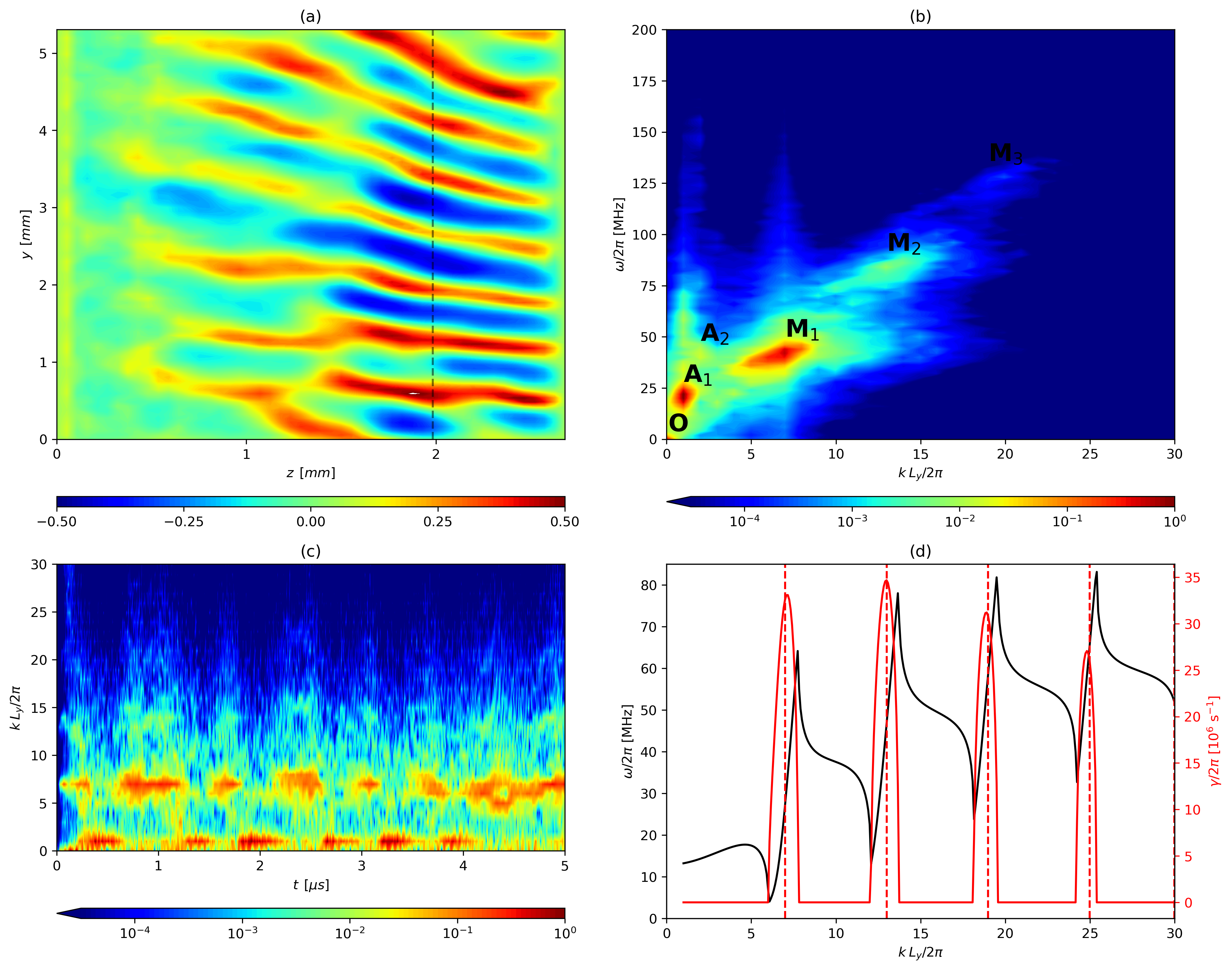}
    \caption{Normalized oscillations of $\phi$ present in the full-PIC simulation. (a) snapshot at $t=2.45\mu s$; (b) normalized PSD dispersion in the $\omega$--$k_{}$ plane; (c) normalized PSD $t$--$k_{}$  spectrogram; (d) Real frequency and growth rate, respectively in black and red, as computed from linear theory \cite{bell24}.  The vertical dashed red lines mark $k_{}L_y/2\pi=7,13,19$ . The dotted black line in (a) denotes the axial slice ($z=1.98mm$) where both the dispersion diagram and spectrogram are computed.}
    \label{fig:data_overview}
\end{figure}

In previous work \cite{bell24}, unstable short-wavelength modes were reported to grow from the initially homogeneous plasma, mainly in the downstream part of the domain, quickly evolving into a nonlinear stage and then saturating.
Taking the perturbation electrostatic potential $\phi$ as a representative variable of the plasma behaviour, a snapshot of it is shown on the upper left panel of Figure \ref{fig:data_overview}.
The analysis in this work focuses mainly on data from the axial slice shown in Figure \ref{fig:data_overview}(a), unless otherwise noted. This slice was chosen because it corresponds to a position where oscillations display a large magnitude. Because all spectra will be taken in this slice, herein we use $k$ to refer to the wavenumber in the azimuthal direction. Additionally, the initial transient stage of the simulation is disregarded, and we keep times from 1$\mu s$ to 5$\mu s$ in which the instability has reached nonlinear saturation.
Unless otherwise noted, all spectral analysis in time is done taking windows of  0.5$\mu s$ with 50\% overlap.

The upper right panel of Figure \ref{fig:data_overview} shows the spectral power distribution of $\phi$ in the ($\omega/2\pi$, $k_{}/2\pi$) plane, revealing two approximately straight dispersion branches, though obscured by additional spectral features. Along these branches, five prominent peaks appear at coordinates (20, 1), (40, 2), (42, 7), (84, 13), and (126, 19), in MHz and multiples of $1/L_y$, respectively. These peaks are designated modes $A_1$ through $M_3$ along what we call the $A$-branch and $M$-branch. The branches correspond with phase velocities of approximately 107 km/s and 32 km/s, respectively 4.5 and 1.3 times the ion sound speed computed in Table \ref{tab:params}. An additional peak, mode $O$, is identified at (0, 1.6). Broadening is observed around modes $M_1$ and $M_2$. Note that the high value of the frequencies (compared to those typical of Hall thrusters) result from using hydrogen ions in the simulation instead of the usual propellants.

Figure \ref{fig:data_overview}(c) showcases the spectrogram of $\phi$, i.e.,
the time evolution of the azimuthal modes.
Plasma properties and the anomalous current  oscillate in time, alternating between periods of growth and quenching, modulated by a low frequency compatible with mode O. The axial character of mode O and the similarity, from Table \ref{tab:params}, of $u_{zi}/L_{z}=0.9 $ MHz
to the mode O frequency, hints at a possible relation with the transit time of ions in the domain.  
The maxima of the amplitude of the ECDI modes in the $M$-branch alternate in time with that of the $A$-branch, sharing the same modulation as the current.

According to ECDI linear theory, modes $M_1$ to $M_3$ are predicted in terms of wavenumber (Figure \ref{fig:data_overview}(d)), while modes $A_1$ and $A_2$, corresponding to the largest azimuthal mode and its harmonic, are not. However, the frequencies of the ECDI modes deviate substantially from linear theory predictions to align with an acoustic-like dispersion. Deviation from linear theory predictions is expected once nonlinear saturation and energy cascading mechanisms disrupt the assumed coherent modes and Maxwellian distributions \cite{janh18}. The dispersion map in Figure \ref{fig:data_overview} further reveals that, although mode $M_2$ is predicted to have the highest growth rate, mode $M_1$ rapidly dominates the spectrum in energy, offering a first hint that nonlinear energy exchanges among modes are a central feature of the discharge dynamics. Despite the peak broadening, discrete modes persist, unlike the continuous dispersion reported in other studies of the ECDI's later evolution \cite{cava13, tsik15, lafl16a}.

\subsection{Anomalous axial transport}
\label{sec:analysis}
\begin{figure}
    \centering
    \includegraphics[width=0.7\textwidth]{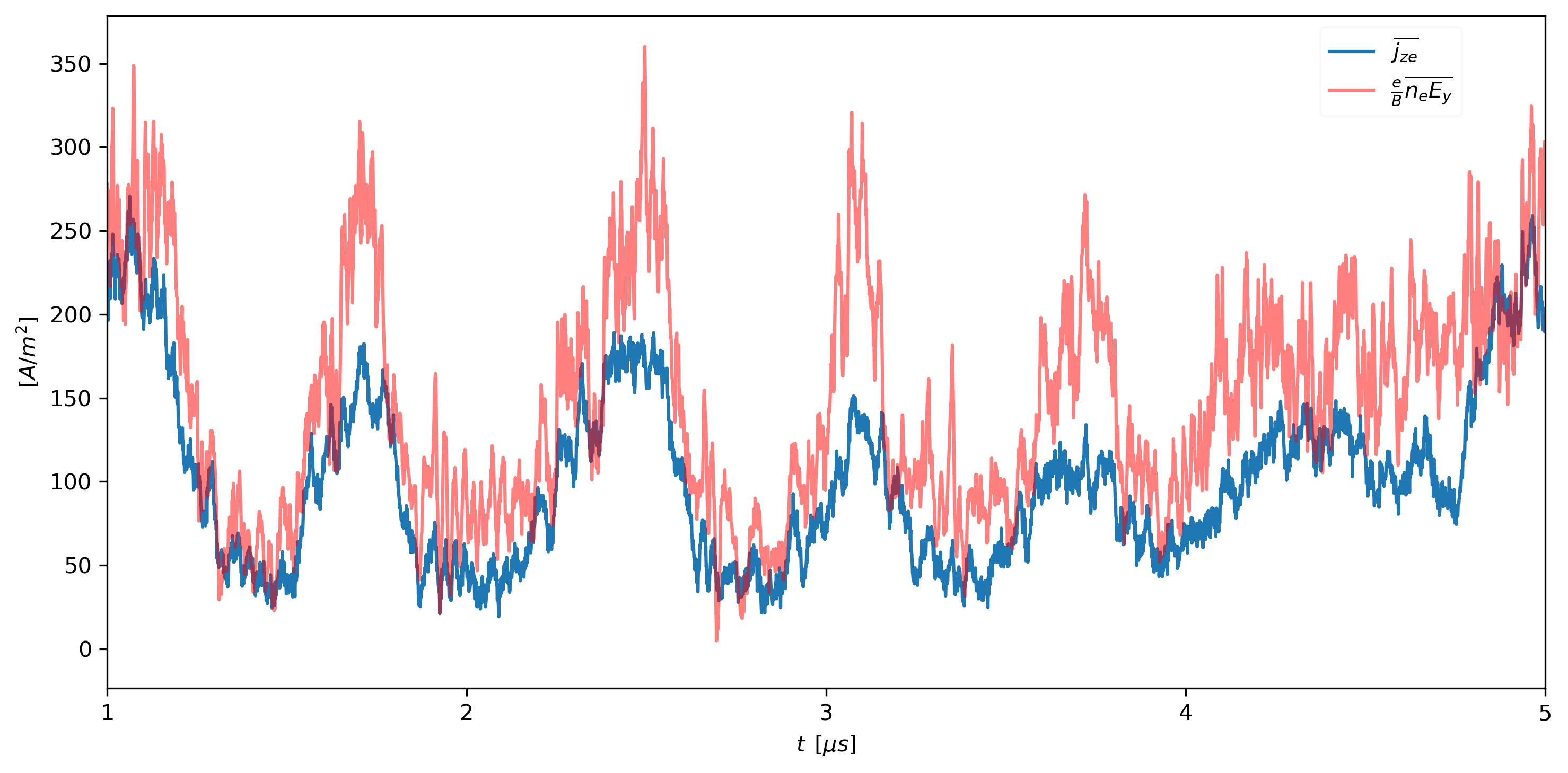}
    \caption{Time-evolution of the electron axial current compared to the current induced by the $n_e$, $E_y$ fluctuations, both measured at $z=1.98mm$ and $y$-averaged ($k_{}L_y=0$).}
    \label{fig:jze_time}
\end{figure}

In the previous section, we identified several oscillation modes are present in the plasma. We now aim to determine how each mode affects cross-field transport. The azimuthal component of the collisionless electron momentum equation at a specific axial position reads
\begin{equation}
    j_{ze}(t, y)=\frac{e}{B}n_eE_y -\frac{m_e}{B} \left ( \frac{\partial }{\partial z}M_{zy e}-\frac{\partial }{\partial y}M_{yy e}-\frac{\partial }{\partial t} n_eu_{ey}\right )
    = j_{ze}^{fluc} + O(m_e/B),
\label{eq:momeq}
\end{equation}
where $M_{zy e}$  and  $M_{yy e}$  are the axial-azimuthal and azimuthal components of the electron momentum tensor, $M_e=\int\int \bm v_e \bm v_e f_ed^3\bm v_e$ , and radial dynamics have been disregarded. The term involving the azimuthal electric field and electron density has been specifically denoted
\begin{equation}
    j_{ze}^{fluc}(t, y)\equiv\frac{e}{B}n_eE_y
\label{eq:jzefluc}
\end{equation}
and it typically dominates transport compared to the other modes, which are $O(m_e/B)$. Indeed, this dominance is 
confirmed in our data, as shown in figure \ref{fig:jze_time}. 

\begin{figure}
    \centering
    \includegraphics[width=1\textwidth]{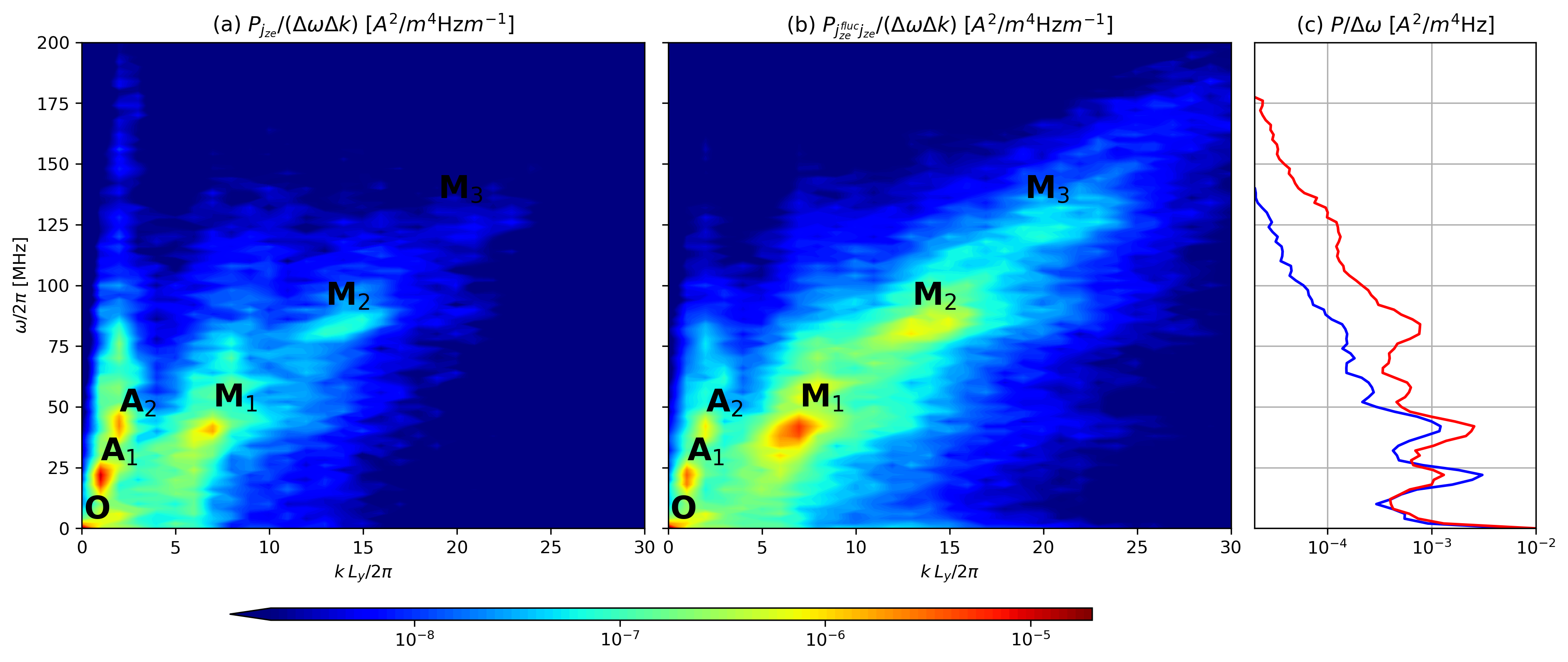}
    \caption{
    Power spectral density of (a) $j_{ze}$ and (b) $j_{ze}^{fluc}$ as taken from the axial slice at $z=1.98mm$. Plot (c) showcases the resulting power from integrating over $k$, with the power of $j_{ze}$ in blue and that of $j_{ze}^{fluc}$ in red. Modes  $O$, $A_1$, $A_2$, $M_1$, $M_2$, $M_3$ correspond to the peaks identified in Figure \ref{fig:data_overview}.}
    \label{fig:jze_spectra}
\end{figure}

In Figure \ref{fig:jze_spectra}, both $P_{j_{ze}}$ and $P_{j^{fluc}_{ze}}$
(i.e., the power spectral density of the left and right hand sides of \eqref{eq:momeq}, once the minor terms are neglected)
are shown for comparison. It is evident that nearly identical modes are present in both and that these are the same modes as in $\phi$, although with varying power levels.
A net axial current forms when the fluctuations in $n_e$ and $E_y$ are in phase; simultaneously, fluctuations in these quantities can be linked to plasma instabilities. 
Notably, however  $j_{ze}^{fluc}$ features a larger power in the ECDI-associated $M$-branch than $j_{ze}$, and greater spectral broadening of these modes. These differences, which are more prominent at larger $k_{}$ and $\omega$, are the neglected terms in equation \eqref{eq:momeq}.

While the full spectrum of $j_{ze}$ is key to understanding collisionless axial transport, practical applications typically focus on the time-averaged (DC) component, often referred to as anomalous transport. 
Figures \ref{fig:jze_time} and \ref{fig:jze_spectra}(c) show that, despite large oscillations in $j_{ze}$ at approximately $20$ and $42$ MHz
, the near-DC transport around $\omega = 0$ MHz is significantly bigger. 
Nevertheless, it is worth noting that AC components of $j_{ze}$ may be rectified in the presence of axial boundary conditions (e.g. an anode surface), potentially resulting in an additional contribution to DC transport. Indeed, boundary conditions can act as an additional source for nonlinear couplings, indicating that the AC part of the spectrum should also be considered in anomalous transport analyses in general.

\begin{figure}
    \centering
    \includegraphics[width=0.75\textwidth]{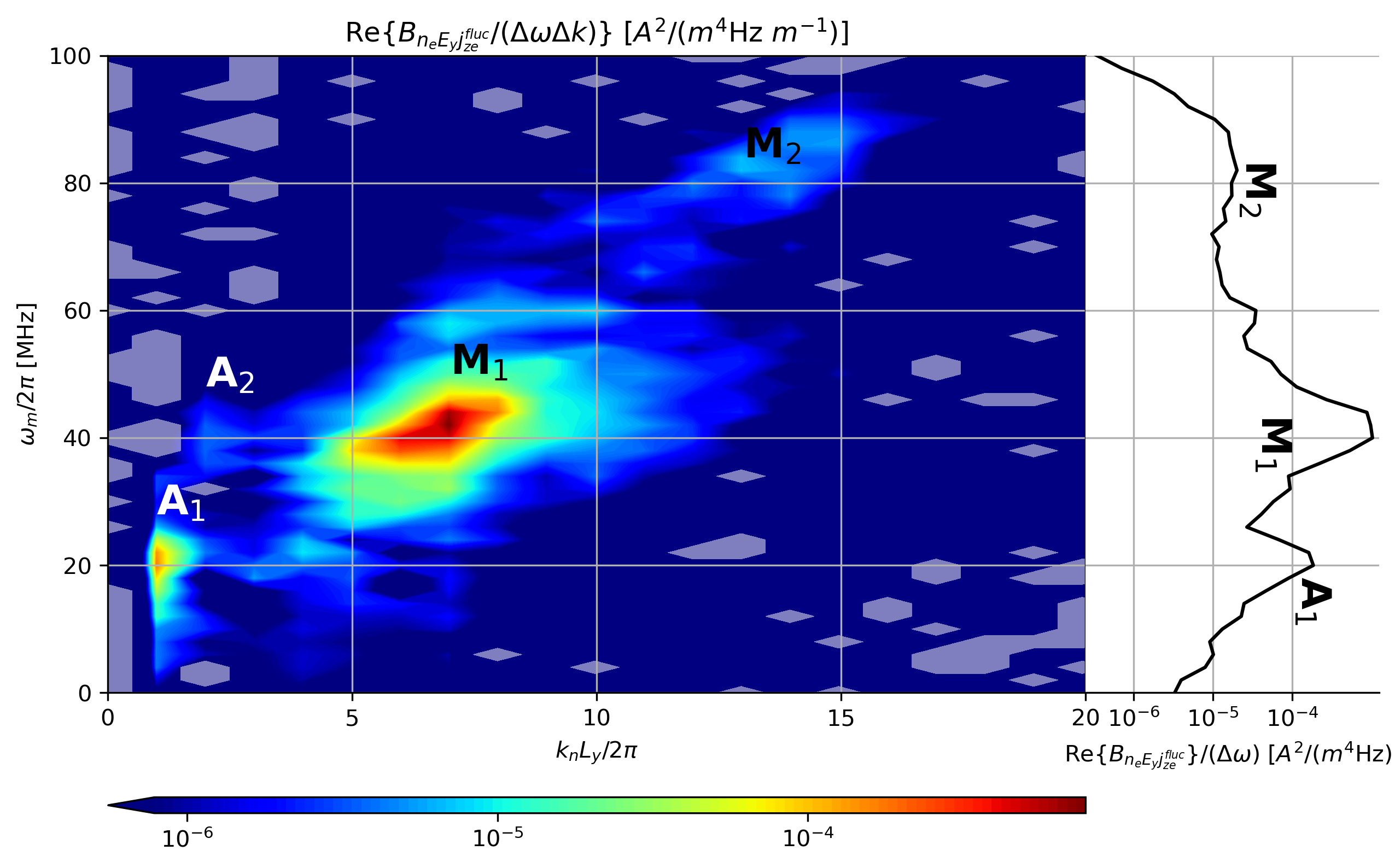}
    \caption{Real part of the cross-bispectrum of $n_e$, $E_{y}$ and $j_{ze}^{fluc}$ , reflecting the spectral contribution to the net cross-field current driven by instability, ($\omega$, $k_{}L_y$) = (0 MHz, 0). Areas shaded in white denote negative values. The plot on the right showcases the resulting power from integrating over $k_n$. Modes $A_1$, $A_2$, $M_1$, $M_2$ correspond to the peaks identified in Figure \ref{fig:data_overview}.}
    \label{fig:jze_contributions}
\end{figure}

So far we have compared the spectra of $j_{ze}$ and $j_{ze}^{fluc}$. However, to univocally link the spectra of the latter as a result of the quadratic phase coupling between modes of $n_e$ and $E_y$ we use the convolution properties of the discrete Fourier transform to write 
\begin{equation}
     \hat{\tilde{\jmath}}^{fluc}_{ze}(\omega, k_{})=\frac{e}{B}\sum_{m}\sum_{n}\hat{\tilde{n}}_e(\omega_m, k_n)\hat{\tilde{E}}_y(\omega-\omega_m, k_{}-k_n).
\end{equation}
Multiplying both sides by $\hat{\tilde{\jmath}}^{fluc\ast }_{ze}(\omega, k_{})$ and averaging over multiple realizations we find
\begin{equation}
    P_{j_{ze}^{fluc}}(\omega, k_{})=\frac{e}{B}\sum_{m}\sum_{n}
    \mathfrak{B}_{n_eE_yj^{fluc}_{ze}}(\omega_m, \omega - \omega_m, k_n, k_{} - k_n)
\end{equation}
This equation makes manifest that the power in $P_{j_{ze}^{fluc}}$ for a particular mode $(\omega, k_{})$ stems from all the quadratic phase coupling between $n_e$ and $E_y$ contained in a slice of the four-dimensional object $\mathfrak{B}_{n_eE_yj^{fluc}_{ze}}$. Focusing on the ($\omega/2\pi$, $k_{}/2\pi$)=(0MHz, 0) component, Figure \ref{fig:jze_contributions} displays the corresponding contributors, with each point corresponding to the coupling of a mode of the electron density $n_e$ at $(\omega_m,k_n)$ and another of $E_y$ at $(-\omega_m, -k_n)$.
On the right of Figure \ref{fig:jze_contributions} a comparison of the $k_{}$-integrated spectral contributors is shown.

The peaks corresponding to modes $A_1$, $M_1$ and $M_2$ are easily identifiable. The magnitude of the bispectrum is most significant for modes $M_1$ and $A_1$, with an order of magnitude difference between their peaks. This conclusion is in disagreement with other studies, which attribute the dominant contribution to the long-wavelength mode \cite{park10, janh18}. Significant broadening in k-space around $\omega/2\pi=42$MHz and between the coherent modes is also observed. Other modes considered, including the remaining ECDI modes, do not yield such significant contributions, and the range above 100 MHz (not shown in the figure) does not seem to feature any relevant quadratic phase coupling. 

\subsection{Three-wave power coupling}
\label{sec:modeling}
 
\begin{figure}
    \centering
    \includegraphics[width=1\textwidth]{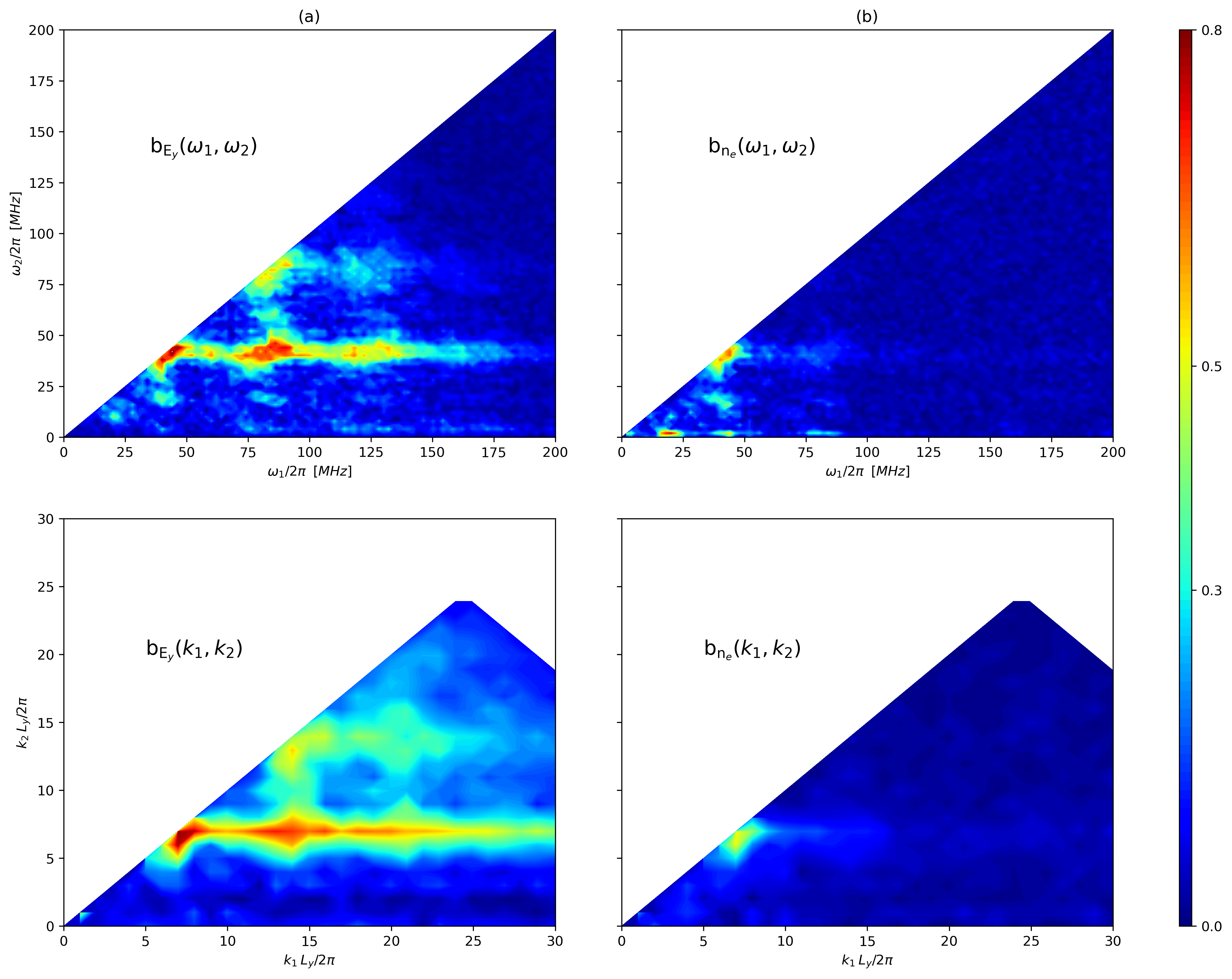}
    \caption{ a) Self-bicoherence of $E_{y}$, b) Self-bicoherence of $n_e$. White space denotes the redundant or inaccessible areas. Upper plots are in $\omega$; lower plots in $k_{}$.}
    \label{fig:bicoherence}
\end{figure}

Linear ECDI theory does not explain the spectrum at saturation, as it would suggest the dominance of  $M_2$ due to its higher linear growth rate. 
This section focuses on elucidating nonlinear power transfers that may exist among the modes in the saturated regime, from a weak-turbulence framework \cite{sagd69,weil77}. 
This is relevant for understanding which are the actual sources of spectral energy at long times (i.e., instabilites in the nonlinear regime) and the channels through which this energy eventually ends up in the modes driving anomalous transport, in particular $M_1$. 

The self-bicoherence of the individual variables helps identify quadratic couplings responsible for nonlinear power transfer, which shape the saturated spectrum. 
Figure \ref{fig:bicoherence}(a) showcases the self-bicoherence for $E_y$  in the  $\omega$ and $k_{}$ domains, below their Nyquist limits.
As the power spectrum in figure \ref{fig:data_overview}(b) is essentially concentrated along a diagonal line, the plots in 
$\omega$ and $k_{}$ present roughly the same features. This permits us, in the following, to refer equivalently to the bicoherence properties in $\omega$ and $k_{}$ for each of the variables. 

In $b_{E_{y}}$, a robust structure,linked to the interaction of ECDI modes $M_1$, $M_2$, $M_3$ and even higher modes, is observed. Specifically, mode $M_1$ has coherent continuous interaction with higher modes, as evidenced by horizontal segment at $\omega_2=42MHz$ with large bicoherence (respectively, $k_{}L_y=7$). We note that the resonant interaction between $M_1$, $M_2$ and $M_3$ must be understood in an broad sense,
because the resonance conditions \eqref{eq:resonanceconditions} are not strictly satisfied among the central $(f,k)$ values of each mode, but only approximately. Indeed, this could be a cause of the observed broadening that the modes exhibit, and may suggest that the interaction involves various neighboring frequencies and wavenumbers, with a power transfer that undergoes various complex three-wave-coupling exchanges among them.
Significantly weaker interactions are seen involving the $A$-branch modes, although relatively small peaks are present corresponding to the triad \triad{A_1A_1}{A_2} and between 
$A_1$ and $A_2$ to their sum-frequency 60 MHz. 
Overall, the high values of the bicoherence (up to $b=0.8$) give us confidence in that wave-wave coupling is a main driver of the late, nonlinear saturated behaviour of the discharge. 

While a large bicoherence is a clear indicator of nonlinear power transfer among modes, an inherent limitation of this spectral quantity is that it does not discriminate by itself the direction in which the energy flows.
To assess this particular aspect, it is necessary to reconstruct a model of the three-wave coupling equations. The basic form of these gives the evolution of the complex wave amplitude\cite{sagd69} of each mode $i$ as
\begin{equation}
    \frac{\partial }{\partial t}{\tilde{x}}_i=(\gamma_i - \text{i} \omega_i) {{\tilde{x}}_i}-v_{gzi}{\frac{\partial }{\partial z}{\tilde{x}}_i}+\sum_{j ,k}V_{ijk}{{\tilde{x}}_j{\tilde{x}}_k},
\label{eq:twc}
\end{equation}
The coefficients on the right-hand side are the linear term with growth rate $\gamma_i$ and frequency $\omega_i$, the $z$ component of the group velocity $v_{gzi}$, and the sum of wave-wave interactions with coupling coefficients $V_{ijk}$, for each mode. The sum of the latter extends to all values of $j,k$ (positive or negative) that yield a valid resonant triad with mode $i$. 
Multiplying equation \eqref{eq:twc} by $\tilde{x}_i^*$ gives, after some algebra
\begin{align}
    \frac{\partial }{\partial t}|\tilde x_i|^2=2\gamma_i 
    |\tilde x_i|^2-v_{gzi}\frac{\partial }{\partial z}|\tilde x_i|^2+\sum_{j ,k}2V_{ijk} \cos\alpha_{ijk} |\tilde x_i \tilde x_j \tilde x_k|,
    \label{eq:twcsquared}
\end{align}
where $\alpha_{ijk}$ is the mutual phase angle resulting from the phases of modes $i,j,k$.

Determining the model coefficients by data regression methods faces two main challenges: the algebraic system in Equation \eqref{eq:twc} is typically overdetermined when considering multiple time instants and/or realizations, and statistical stationarity of the signals is often required for applicability. However, strict stationarity also renders $\partial \hat{\tilde{x}}_i/\partial t$ negligible compared to the terms on the right-hand side, making the system, at the same time, ill-posed (i.e., if the left hand side for all equations were strictly zero, coefficients can be determined up to a multiplying factor). This latter issue is not usually acknowledged. 
The prevailing regression schemes are due to Kim and Ritz  \cite{kim96, ritz89} and De Wit  \cite{wit99}. In essence, they obtain a single fit for the $\gamma_i$, $\omega_i$, $v_{gzi}$ vectors and the $V_{ijk}$ matrix simultaneously from several higher-order spectra; as a consequence, these methods are usually computationally expensive. Furthermore, they typically result in full matrices with an enormous list of terms on each mode's equation, which complicates interpretability: other than the magnitude of the coefficients, these methods do not discriminate which
terms in each mode's equation are more fundamental and which more accessory. That is, we have no clear indication of how the fit error would deteriorate, were we to drop one particular term in the equation. This question is relevant when we intend to derive simple, understandable models that capture the dominant nonlinear dynamics.

Here, we  instead use the SINDy algorithm to identify a hierarchy of reduced models for the dominant modes only
($M_1$, $M_2$ and $M_3$) of $E_y$ at $z=1.98mm$. While this does not remove the ill-posedness discussed above, 
it allows us to effectively order the right hand side terms according to their significance in the dynamics, and to truncate the hierarchy at a sensible point that balances model accuracy and complexity.
One may wonder to what extent such a model is complete, given the exclusion of the $A$-branch and mode $O$ from the modeling. However, it has been shown that their contribution to transport is minimal compared to that of mode $M_1$, and separate in time in the case of the $A$-branch, while bicoherence analysis demonstrates the coupling of these modes to the $M$ dispersion branch is minimal, too. Thus, they can interpreted as independent processes/instabilities, and can be modeled as such, focusing here solely on the main responsible for transport.

First, and reasoning from equation \eqref{eq:twcsquared},
we take the ansatz that the 
mode interactions remain moderately coherent over time \cite{sagd69} such that $\cos\alpha_{ijk}$ in 
equation \eqref{eq:twcsquared}
remains essentially constant. This assumption is supported by the large bicoherence values found among the modes, which would diminish if the mutual phase were not stable.
Next, 
to take into account mode broadening, we select a $(\omega,k)$ rectangle of $\Delta\omega/2\pi=0.7$MHz, $\Delta kL_y/2\pi=1$ around modes $M_1$, $M_2$, $M_3$ and average their PSD $P_i$ ($i=1,2,3$)
over them. This $P_i$ is computed on time windows to obtain a time signal that can be used in the SINDy algorithm. The averaging and the time windowing has the additional advantage of reducing the susceptibility of the results to noise, as in the weak-SINDy variant of the technique \cite{mess21a} .
With these premises, we can expect $P_i$ to follow an equation analogous to
\begin{align}
    \frac{\partial }{\partial t}P_i \simeq 
    2\gamma_i {P_i}-v_{gzi}{\frac{\partial }{\partial z}P_i}+\sum_{j ,k}2V_{ijk}^\prime {\sqrt{P_iP_jP_k}}.
\label{eq:powertwc}
\end{align}

We choose a window spanning 1.4$\mu s$ with a step of 0.3$\mu s$. An optimal range for step size and window size balances capturing slow amplitude variations and maintaining spectral resolution. Too small a step size amplifies noise, while too large misses key variations. Similarly, improper window sizing can mix mode energies or fail to represent broadened interactions. Sensitivity analysis within a  $\pm 0.1\mu s$  range confirmed the stability of coefficients and conclusions.

We are then able to apply the SINDy algorithm to separately identify the evolution equations for the magnitude of each mode. For the three modes considered, two triads are possible: \triad{M_1M_1}{M_2} and \triad{M_1M_2}{M_3}. Note that the resonance condition also accounts for negative frequencies; for example, for the frequency triplet $\omega_1\approx84$ MHz, $\omega_2\approx-42$ MHz, $\omega_3\approx42$ MHz can be seen as mode $M_1$ interacting with mode $M_2$ to either transfer or receive energy, depending on the sign of the coefficient. 
All possible terms in the right hand side of equation \eqref{eq:powertwc} are considered in the feature library of the search algorithm.
Finite differences are used to compute spatial and temporal derivatives of $P_i$.

\begin{table}[]
    \centering
    \begin{tabular}{|c|c|p{14cm}|}
    \hline
         & \boldsymbol{$R^2$} & \textbf{Model} \\
        \hline
    1 & 0.01& $\partial P_1/\partial t = 1.25\cdot10^{-3} \textcolor{blue}{\sqrt{P_1P_1P_2}}$ \\
    \textbf{2} & \textbf{0.12}& \boldsymbol{$\partial P_1/\partial t=- 8.72\cdot10^{5} \textcolor{red}{P_1} + 8.90\cdot10^{-3} \textcolor{blue}{\sqrt{P_1P_1P_2}}$} \\
    3 & 0.13& $\partial P_1/\partial t=- 9.2\cdot10^{5} \textcolor{red}{P_1} + 9.06\cdot10^{-3} \textcolor{blue}{\sqrt{P_1P_1P_2}} - 9.10\cdot10^{1} \textcolor{orange}{\partial P_1/\partial z}$ \\
    4 & 0.13& $\partial P_1/\partial t=- 8.22\cdot10^{5} \textcolor{red}{P_1} + 6.40\cdot10^{-3} \textcolor{blue}{\sqrt{P_1P_1P_2}} + 4.02\cdot10^{-2} \textcolor{cyan}{\sqrt{P_1P_2P_3}} - 1.14\cdot10^{2} \textcolor{orange}{\partial P_1/\partial z}$ \\
    \hline\hline
    1 & 0.03& $\partial P_2/\partial t= -6.18\cdot10^{-5} \textcolor{blue}{\sqrt{P_1P_1P_2}}$ \\
    \textbf{2} & \textbf{0.07}& \boldsymbol{$\partial P_2/\partial t= 5.94\cdot10^{5} \textcolor{red}{P_2} - 1.66\cdot10^{-4}\textcolor{blue}{\sqrt{P_1P_1P_2}}$} \\
    3 & 0.07& $\partial P_2/\partial t= 6.20\cdot10^{5} \textcolor{red}{P_2} - 1.63\cdot10^{-4}\textcolor{blue}{\sqrt{P_1P_1P_2}} - 2.68\cdot10^{-3}\textcolor{cyan}{\sqrt{P_1P_2P_3}}$ \\
    4 & 0.07& $\partial P_2/\partial t= 6.34\cdot10^{5} \textcolor{red}{P_2} - 1.66\cdot10^{-4}\textcolor{blue}{\sqrt{P_1P_1P_2}} - 2.48\cdot10^{-3}\textcolor{cyan}{\sqrt{P_1P_2P_3}} - 1.01\cdot10^{1}\textcolor{orange}{\partial P_2/\partial z}$ \\
    \hline\hline
    1 & 0.01& $\partial P_3/\partial t=  - 4.72\cdot10^{-5} \textcolor{cyan}{\sqrt{P_1P_2P_3}}$ \\
    \textbf{2} & \textbf{0.04}& \boldsymbol{$\partial P_3/\partial t= 7.00\cdot10^{5} \textcolor{red}{P_3} - 1.89\cdot10^{-4} \textcolor{cyan}{\sqrt{P_1P_2P_3}}$} \\
    3 & 0.06& $\partial P_3/\partial t= 6.62\cdot10^{5} \textcolor{red}{P_3} - 1.84\cdot10^{-4} \textcolor{cyan}{\sqrt{P_1P_2P_3}} - 4.64\cdot10^{1} \textcolor{orange}{\partial P_3/\partial z}$ \\
    \hline
    \end{tabular}
    \caption{Pareto front models for modes $M_1$, $M_2$, $M_3$, obtained by sparse regression. $P_i$ denotes the power spectral density of mode $i$. Terms in \textcolor{red}{red} are associated to linear growth/decay rates; \textcolor{orange}{orange} designates convection terms; \textcolor{blue}{blue} and \textcolor{cyan}{cyan} terms highlight related wave-wave couplings terms across the modes. The Pareto-optimal models are highlighted in \textbf{bold}.}
    \label{tab:models}
\end{table}

A Pareto front, or hierarchy, of increasingly more complex models for each mode is obtained and displayed in table \ref{tab:models}. In bold we have highlighted our identified Pareto knee model for each mode, which corresponds to the biggest increase in score with the addition of a single term. The relative importance of a term in its respective equation is represented by how early it appears in the hierarchy. Note that once the $R^2$ score has saturated (i.e., it does not increase significantly with the addition of new terms) experience dictates a loss of physical meaningfulness, as the term added is probably linked to an overfit.

Although the  $R^2$ score stabilizes beyond the Pareto knee model (except perhaps for $M_3$), suggesting reliable model identification \cite{hans92}, the maximal score of this simplified model is low (ranging from $0.03$ to $0.13$ as per Equation \eqref{eq:R2score}). This reveals that our basic model is likely too simple, and that additional terms (such as interactions with other modes) that have been neglected may play an important role. 
Nevertheless, we shall assume that the weak fit obtained  correctly captures the essential dynamics among the retained modes.
Thus, we focus on the order of appearance, the sign and magnitude of the $\gamma_i$ and $V_{ijk}$ coefficients, where positive/negative signs indicate energy flow into/out of the mode, respectively.

In the models of $M_1$ and $M_2$, the dominant term (and therefore the first that shows up in the hierarchy) is the coupling term for the triad \triad{M_1M_1}{M_2}. This coupling transfers energy from mode $M_2$ into mode $M_1$ according to the sign structure of the coefficients.
The term coupling modes \triad{M_1M_2}{M_3} appears first in the model for mode $M_3$, and also shows up for models more complex than the Pareto-optimal ones for $M_1$ and $M_2$. This term would convey energy from $M_3$ into $M_2$ and $M_1$.
The term that appears second in all three modes is the growth/decay rate, and its inclusion suffices to reach the defined Pareto optimality.
The growth rate signs indicate power input into modes $M_2$ and $M_3$ and dissipation in mode $M_1$.
Finally, axial convection,  at least at this position ($z= 1.98$ mm) plays an overall minor role, as the corresponding terms do not show up until after the optimal model in the hierarchy. This is not surprising, as we are analyzing an axial position that displays little axial variability of the spectrum.

Regarding the magnitude of the terms, we note that if these three modes were an isolated system, we would expect exact energy conservation, with the coefficients for each triad in each equation adding up to zero. This is clearly not so, as they differ in at least one order of magnitude for each of the two nonlinear couplings recovered in the model.
Apart from the error of the fit, to which part of this incongruence is perhaps attributable, we acknowledge that this reduced model does not cover for all the nonlinear power couplings likely present in the data, and that a model with energy conservation would need to  take into account all possible interactions, a task beyond the scope of this work. 

The magnitudes of the growth rates $\gamma_i$ are comparable among the three 
modes and two orders of magnitude lower than predicted by linear theory (Figure \ref{fig:data_overview}), consistent with the plasma being in a saturated state (similar decrease in the growth rate with respect to linear theory are have been experimentally \cite{brow23, tsik09b}).

\begin{figure}
    \centering
    \includegraphics[width=0.6\textwidth]{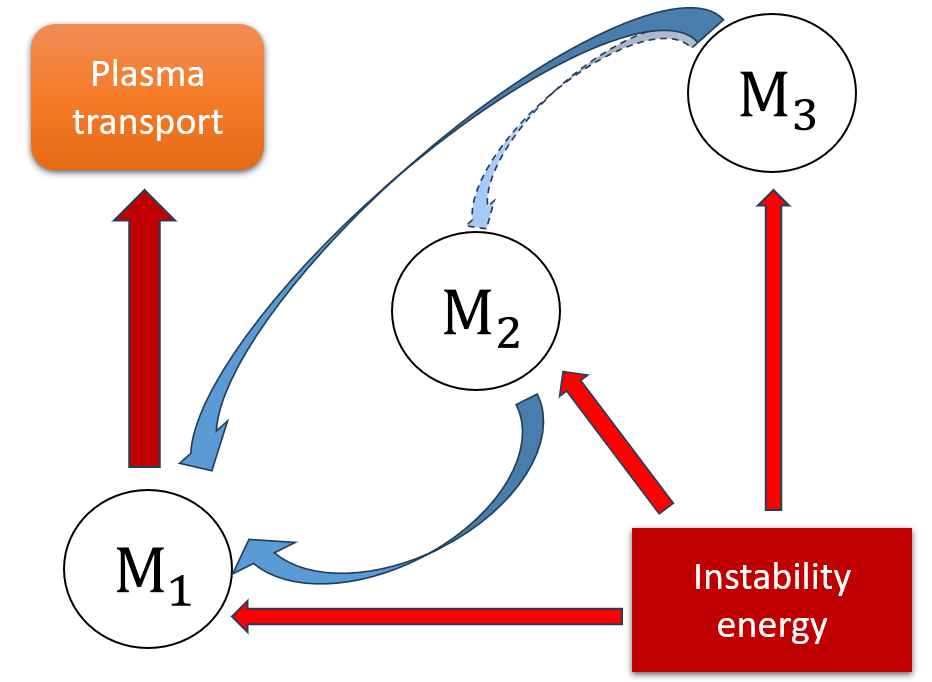}
    \caption{Schematic of the power flux in the Pareto-optimal reconstructed  model in table \ref{tab:models}, with arrows representing energy transfer. The dashed arrow implies uncertainty.}
    \label{fig:modeloutcome}
\end{figure}

From the analysis of the linear growth rate and the quadratic coefficients, a general picture emerges: the sign structure of the couplings \triad{M_1M_1}{M_2} and \triad{M_1M_2}{M_3} suggest an inverse energy cascade, where energy of the higher-frequency, higher-k modes flows toward lower-frequency, lower-k modes. Modes $M_2$ and $M_3$ receive energy from the ECDI instability, which is then transferred to the longer-scale mode $M_1$. The negative growth rate of mode $M_1$ can be attributed to a net energy loss between what it gains from the instability at all modes and what it loses to generating anomalous plasma transport. 
This energy flow has been schematically represented in Figure \ref{fig:modeloutcome}.

Finally, we return to figure \ref{fig:bicoherence} to discuss the structure of the self-bicoherence of $n_e$. In comparison to that of $E_y$, $n_e$ presents significant differences.
Indeed, another kind of interactions appears to dominate for $n_e$, in particular the modulation of modes $A_1$ and $M_1$ by mode $O$. The next strongest interaction is still the triad \triad{M_1M_1}{M_2}, but the rest of the $M$-branch displays significantly lower bicoherence than in the case of $E_y$. The substantial differences among the two variables hint at the possibility that a two-field description of the nonlinear dynamics is necessary, that better represents the energy flux between the modes of the field (in $E_y$) and the particles (in $n_e$). This additional complexity may well be necessary as well to raise the score values in the models obtained here.

\section{Conclusions}
\label{sec:conclusions}

A higher-order spectral analysis of a 2D-$E\times B$ kinetic simulation was conducted. It identified two distinct dispersion branches within the ion-acoustic range, and observed that a low-frequency mode modulated the discharge. Anomalous transport was seen to be primarily driven by in-phase fluctuations in density and electric field arising from the longest-wavelength mode of the Electron Cyclotron Drift Instability (ECDI), $M_1$. Other modes like $A_1$ and $M_2$ contribute as well. The axial electron current is seen to be modulated by mode $O$, and the $M$ and $A$ modes appear to alternate in time accordingly.

Our bicoherence analysis revealed nonlinear power transfer across scales in the $E_y$ spectrum, and sparse regression (SINDy) was employed to develop a hierarchy of data-driven spectral models based on three-wave coupling equations for ECDI modes. These methods indicate these wave-wave interactions are a key energy exchange mechanism, and that the most likely structure of this energy flow is an inverse energy cascade. Nevetheless, the low fit scores cast a reasonable doubt on this last conclusion, and hint at the existence of other wave-wave couplings with other modes that need to be retained for a fuller picture.

The capability to quantitatively separate different mode contributions to plasma transport appears crucial for studies with co-existing instabilities, such as ECDI combined with lower hybrid or modified two stream instabilities \cite{janh18, hara20, mike20, brow23}. Additionally, the approach for determining growth rates and nonlinear energy exchange direction enables isolation of the key drivers of dynamics. We note that for a perfectly-stationary spectrum the determination of the coupling coefficients becomes ill-posed, an aspect that affects all regression techniques, including ours. 

This general methodology used may be applicable to other plasma transport studies driven by fluctuations.
Future research could refine present results by incorporating wavelet analysis to better account for temporal variations of broadened structures, and explore models with coupling terms on two fields ($E_y$ and e.g. $n_e$, which show distinct bicoherence diagrams) to gain further insight into the oscillation dynamics.

\section*{Data availability statement}
The data that support the findings of this study are openly available at the following URL/DOI: XXXXXXXX.

\section*{Acknowledgments}

The authors would like to thank Matteo Ripoli and the rest of the EP2 research group at Universidad Carlos III de Madrid, for the invaluable overarching discussions. This project has received funding from the European Research Council (ERC) under the European Union’s Horizon 2020 research and innovation programme (ZARATHUSTRA project, grant agreement No 950466).
Additionally, during this period Bayón-Buján enjoyed a grant from the Consejería de Educación, Universidades, Ciencia y Portavocía of the Community of Madrid (grant PEJ-2021-AI/TIC-23158).

\bibliographystyle{aiaa}
\bibliography{bibtex/ep2,bibtex/others}

\end{document}